\documentclass[conference]{IEEEtran}
\usepackage{amsmath,amsfonts}
\usepackage{algorithmic}
\usepackage{algorithm}
\usepackage{array}
\IEEEoverridecommandlockouts
\usepackage[caption=false,font=normalsize,labelfont=sf,textfont=sf]{subfig}
\usepackage{textcomp}
\usepackage{stfloats}
\usepackage{url}
\usepackage{verbatim}
\usepackage{caption}
\usepackage{graphicx}
\usepackage{cite}
\usepackage[switch]{lineno}
\usepackage{amsmath,amssymb,amsfonts}
\usepackage{orcidlink}
\usepackage{comment}
\usepackage{algorithmic}
\usepackage{textcomp}
\usepackage{tabularx}
\usepackage{titlesec}
\usepackage{graphicx}
\usepackage{subcaption} 
\usepackage{threeparttable}
\usepackage{textcomp}
\usepackage{xcolor}
\usepackage{cite}
\usepackage{enumitem}
\usepackage{makecell}
\usepackage{moresize}
\definecolor{darkgreen}{RGB}{0,100,0}
\usepackage{array} 
\usepackage{hyperref}
\hypersetup{
    colorlinks=true,
    linkcolor=blue,
    citecolor=blue,
    urlcolor=blue
}
\newcommand{\rr}[1]{{\color{black} #1}}
\newcommand{\rrr}[1]{{\color{black} #1}}

\hypersetup{
    hidelinks,
    colorlinks=true,
    citecolor=blue,
    linkcolor=blue,
    urlcolor=blue
}
\def\BibTeX{{\rm B\kern-.05em{\sc i\kern-.025em b}\kern-.08emT\kern-.1667em\lower.7ex\hbox{E}\kern-.125emX}}

\begin{document}
\bstctlcite{IEEEtran:BSTcontrol}
\title{Experimental Characterization of SMD Inductors for Cryogenic Power Conversion at 4~K}

\author{
\IEEEauthorblockN{
Rami~Rasheedi\,\orcidlink{0009-0009-8410-8011}$^{1,2}$,
Umeshkumar~Patel$^{2}$,
Jonathan~Baldwin$^{2}$,\\
Antonino~Miceli\,\orcidlink{0000-0001-5994-5402}$^{2}$,
and Inna~Partin-Vaisband\,\orcidlink{0000-0002-6399-6672}$^{1}$
}
\IEEEauthorblockA{
$^{1}$Department of Electrical and Computer Engineering, University of Illinois Chicago, Chicago, IL, USA\\
$^{2}$Argonne National Laboratory, Lemont, IL, USA
}
}




\maketitle

\begin{abstract}
\rrr{With the increasing complexity and scalability demands of cryogenic electronic systems, such as quantum computing platforms, conversion of power within the cryogenic environment has become a primary concern. Magnetic inductors are among the key components affecting converter performance, as their electrical and magnetic properties can change significantly at low temperatures. In this paper, five commercially available Vishay SMD power inductors are experimentally characterized down to 4~K: DC resistance is measured continuously during cooldown, while frequency-dependent AC inductance and resistance are measured up to 2~MHz at both room temperature and 4~K. The results show substantial DC resistance reductions of 25$\times$--58$\times$ at 4~K, while the inductance degrades by only about 5--5.5\%. Although cryogenic operation slightly increases the AC resistance (due to the intensified skin effect) and raises core losses, converter-level analysis shows an inductor-only efficiency improvement of approximately 5.3--6.9 percentage points at 1~A, and up to 22.8 points at 3~A. The measured frequency-dependent impedance of each inductor is imported into Cadence and a complete single-phase 24~V-to-1~V buck converter with GaN power devices is simulated at 1, 2, and 3~A at 4~K. The full-converter efficiency at 4~K is shown to improve by 3.0--5.2 percentage points at 1~A and by up to 22.5 points at 3~A for the AC-characterized devices. These findings indicate that commercially available off-the-shelf power inductors remain operationally viable under cryogenic conditions and can simultaneously increase inductor efficiency and reduce the overall power loss in cryogenic power conversion systems.}
\end{abstract}

\begin{IEEEkeywords}
Cryogenic Power Delivery, Power Conversion, Inductors, Quantum Computing
\end{IEEEkeywords}

\section{Introduction}
\IEEEPARstart{R}{elatively} low power is typically consumed by quantum circuits operating at cryogenic temperatures. However, a substantial amount of surrounding support electronics, including digital processing units, analog interfaces, readout circuits, and control systems, is required by the scalability of quantum computing platforms, and moderate to high power levels can collectively be demanded by these \cite{patra2020characterization,yoo2023design,guo202429, rasheedi202528}. While portions of the associated electronics may also be placed inside the cryogenic environment to improve scalability and reduce signal routing complexity \cite{berlingard2021rf,charbon2016cryo,patra202019,guo202429, gharib2026cryogenic}, the power conversion and supply circuits supporting these systems are still commonly located outside the cryogenic refrigerator and are operated at room temperature.
As a result, significant conduction losses, thermal loading, and signal degradation can be introduced by long interconnects between the external power delivery circuitry and the cryogenic electronic systems, particularly in applications requiring high current delivery \cite{claassen2005inductor}.
As quantum processors continue to scale toward larger qubit counts, the thermal load, Joule heating, and physical routing complexity associated with the growing number of interconnects are expected to become major bottlenecks for cryogenic system scalability \cite{krinner2019engineering,zou2025power,martin2022energy}.

Consequently, integration of power conversion within the cryogenic environment has emerged as a promising approach to reduce power-delivery losses and improve system scalability  \cite{zou2025power}.
%
\begin{figure}[tb]
\centerline{\includegraphics[width=\columnwidth]{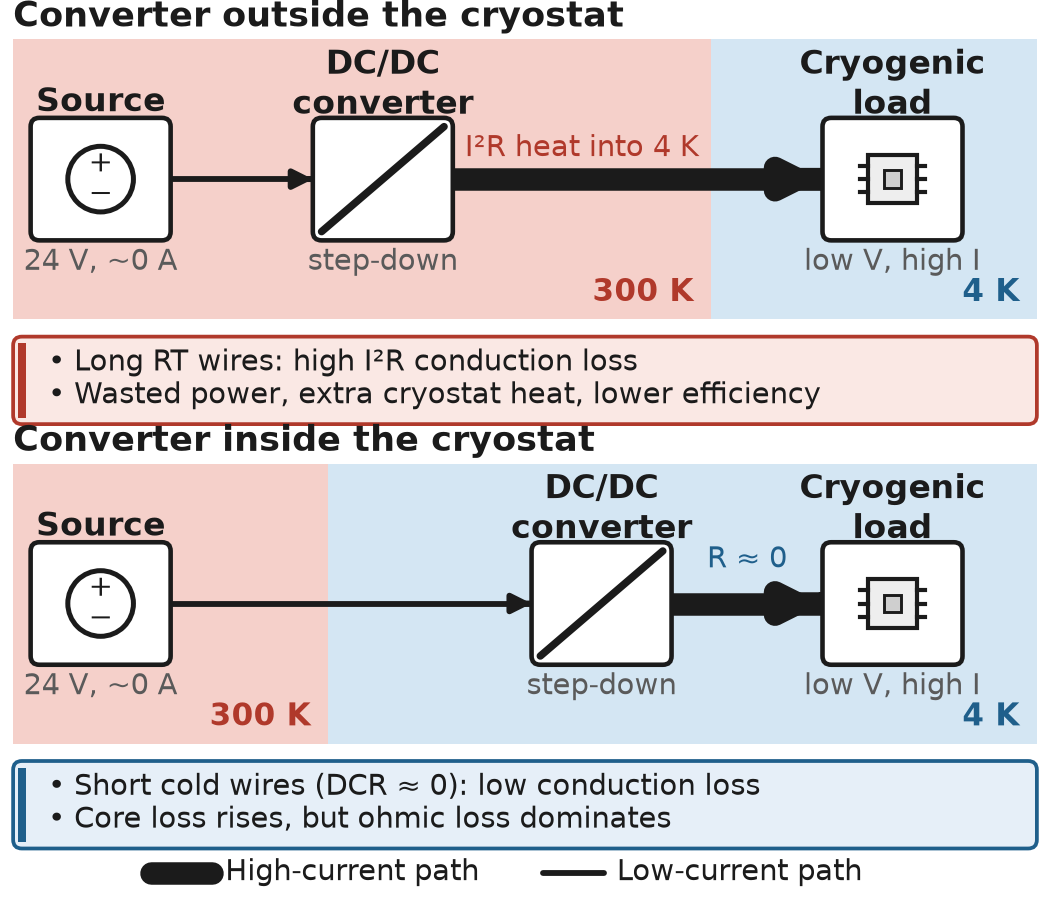}}
\caption{Comparison of room-temperature and cryogenic power converter placement for power delivery to 4~K electronics.}
\label{intro_fig}
\end{figure}
However, the operational behavior and performance trade-offs of power conversion components under cryogenic conditions remain insufficiently understood, and detailed characterization and analysis are required \cite{yin2021characterization}.
Among the various components required for cryogenic power conversion, passive magnetic components remain comparatively less explored than cryogenic CMOS circuits, voltage references, current references, and low-dropout regulators \cite{andrade2021cryogenic,van2024cryo,ray20255}.
Specifically, inductors are considered one of the most critical and loss-dominant elements in high-frequency power converters \cite{rasheedi2026package,rasheedi2026adaptive}. Complex and often difficult-to-predict trade-offs are exhibited by their behavior under cryogenic operation due to the simultaneous variation of electrical and magnetic material properties with temperature. At cryogenic temperatures (e.g., 4~K), the DC resistance of the winding conductors decreases significantly because of the reduction in conductor resistivity, whereby winding conduction losses are reduced and converter efficiency is potentially improved \cite{berlingard2021rf}.
However, degradation in magnetic performance may be simultaneously experienced under cryogenic operation by magnetic-core inductors, which are widely preferred over air-core implementations because of their substantially higher inductance density and compact size \cite{yin2021characterization}. In particular, the electrical resistivity of the magnetic core material can decrease at low temperatures, by which eddy-current-related losses are increased according to
\begin{equation}
P_{ac}^{eddy} = \frac{C B^2 f_{sw}^2 d^2}{\rho},
\end{equation}
where \(B\), \(f_{sw}\), \(C\), \(d\), and \(\rho\) are the peak flux density, switching frequency, proportional constant, lamination thickness, and electrical resistivity of the magnetic material, respectively \cite{buttner2026loss}. 
In addition, the frequency-dependent AC winding resistance can increase more aggressively with frequency under cryogenic operation due to the intensified skin effect caused by the reduced conductor resistivity. The AC resistance of the winding can be approximated by
\begin{equation}
R_{ac} \approx R_{dc}\frac{r}{2}
\sqrt{\frac{\pi f \mu}{\rho_w}}
\label{eq:ac_skin_effect}
\end{equation}
where 
$R_{dc}$ is the DC resistance of the conductor, $r$ is the radius of the conductor, $2\pi f$ is the angular frequency corresponding to the operating frequency $f$, $\mu$ is the magnetic permeability of the conductor medium, and $\rho_w$ is the electrical resistivity of the winding material. While the DC resistance decreases substantially at cryogenic temperatures, the current penetration depth is simultaneously reduced by the reduction in $\rho_w$, whereby the ratio between $R_{ac}$ and $R_{dc}$ is caused to increase more rapidly with frequency compared to room-temperature operation.
\begin{table}[t]
\centering
\caption{Comparison of Non-air Core Inductor Behavior at Room (300~K) and Cryogenic (4~K) Temperatures.}
\label{tab:cryo_inductor_tradeoffs}
\renewcommand{\arraystretch}{1.2}
\footnotesize
\begin{tabular}{l|c|c|l}
\hline
\textbf{Parameter} & \textbf{300~K} & \textbf{4~K} & \textbf{Dominant component} \\ \hline

\textbf{$\mathbf{R_{dc}}$} &
Nominal &
$\downarrow \downarrow \downarrow$ &  winding resistance \\ \hline

\textbf{Core loss} &
Nominal &
$\uparrow$ & eddy-current \\ \hline

\textbf{Inductance ($\mathbf{L}$)} &
Nominal &
$\downarrow$ &  core permeability drop\\ \hline

\textbf{$\mathbf{R_{ac}}$/\textbf{$\mathbf{R_{dc}}$} vs.\ frequency} &
Nominal &
$\uparrow \uparrow \uparrow$ & skin effect \\ \hline

\textbf{Permeability ($\boldsymbol{\mu}$)} &
Nominal & 
$\downarrow$ &  domain wall mobility\\ \hline

\hline

\end{tabular}
\end{table}

It has also been shown by previous studies that the magnetic permeability of certain core materials can degrade substantially at cryogenic temperatures \cite{buttner2026loss}. In particular, ferrite materials have been reported to exhibit permeability reductions exceeding one order of magnitude in some cases, while significantly better cryogenic stability is generally demonstrated by nanocrystalline and powder-core materials \cite{gerber2002performance}. The achievable inductance value can be directly reduced and the magnetic behavior of the converter can be altered by such permeability degradation. In parallel, substantial increases in cryogenic core losses for ferrite materials have been reported by several studies, whereas comparatively smaller degradation tends to be exhibited by amorphous and nanocrystalline materials \cite{claassen2005inductor}.
The setup for the conventional room-temperature converter placement and cryogenic converter integration is shown in Fig.~\ref{intro_fig}, illustrating the reduction of interconnect-related conduction losses at the expense of operating the converter under cryogenic conditions.
The primary trade-offs in non-air-core inductor behavior between room-temperature and 4~K operation are summarized in Table~\ref{tab:cryo_inductor_tradeoffs}.

A complex design trade-off is created by these competing effects, which must be carefully evaluated so that it can be determined whether a net efficiency improvement is ultimately provided by operating power converters at cryogenic temperatures, or whether the benefits obtained from reduced conduction losses are outweighed by the degradation in magnetic component performance. Accordingly, the cryogenic behavior of commercially available off-the-shelf inductors is experimentally investigated in this work so that their suitability, operational validity, and efficiency trade-offs can be evaluated for cryogenic power conversion systems intended for quantum and cryogenic electronic applications.

\rrr{Prior investigations of magnetic components at cryogenic temperatures fall into two categories that do not overlap with the present study. The first is concerned with radio-frequency inductors: on-chip spiral inductors integrated in an FD-SOI technology~\cite{berlingard2021rf} and on-chip microwave passive components~\cite{patra2020characterization} have been characterized down to 4~K for quantum-control integrated circuits, and even commercial SMD inductors have been examined at 4~K, but only as nanohenry-class RF matching elements rather than power components~\cite{volkhin2025experimental}. The second category is concerned with power conversion but is based on bare or in-house-wound magnetic-core samples: soft-magnetic core materials have been characterized for cryogenic use~\cite{yin2021characterization,gerber2002performance,buttner2026loss}, and custom cryogenic inductor designs have been studied for superconducting power electronics~\cite{claassen2005inductor}. Among these, in the study closest to the present work, in-house-wound bare cores are operated in a cryogenic buck converter, yet only down to 77~K and without commercial packaged devices~\cite{yin2021characterization}. To the best of the authors' knowledge, this is therefore the first work in which commercial off-the-shelf SMD power inductors, including their frequency-dependent AC loss, are experimentally characterized and are evaluated at the converter level for DC--DC power conversion down to deep-cryogenic (4~K) temperatures.}

The remainder of this paper is organized as follows. The experimental setup and measurement results for the investigated Vishay SMD power inductors are provided in Section~\ref{Experimental_Measurements}. A performance comparison between converter operation at 4~K and room temperature, along with the corresponding analysis and key observations, is given in Section~\ref{Performance_Comparison}. Finally, the conclusions of this work are summarized in Section~\ref{Conclusion}.

\begin{table}[t]
\centering
\caption{Investigated Off-the-Shelf Inductors.}
\label{tab:inductor_models}
\renewcommand{\arraystretch}{1.2}
\footnotesize
\begin{tabular}{l|c|c|c}
\hline
\textbf{Model Name} & \textbf{Channel} & \textbf{\rrr{Max $R_{dc}$}} & \textbf{\rrr{$L_{dc}$}} \\ \hline

\textbf{IFSC1515AHER470M01} & 2 & 594~m$\Omega$ & \rrr{47~\textmu H} \\ \hline

\textbf{IHLP2020BZER6R8M5A} & 3 & 135.8~m$\Omega$ & \rrr{6.8~\textmu H} \\ \hline

\textbf{IHLP2020BZER5R6M01} & 4 & 122~m$\Omega$ & \rrr{5.6~\textmu H} \\ \hline

\textbf{IHLP1212BZEZ3R3M1Z} & 5 & 54.9~m$\Omega$ & \rrr{3.3~\textmu H}\\ \hline

\textbf{IHLP1212BZEV3R3MA1} & 6 & 98.48~m$\Omega$ &\rrr{3.3~\textmu H}\\ \hline

\end{tabular}
\end{table}

\section{Experimental Measurements}
\label{Experimental_Measurements}

\begin{figure}[h!]
\centerline{\includegraphics[width=0.92\columnwidth]{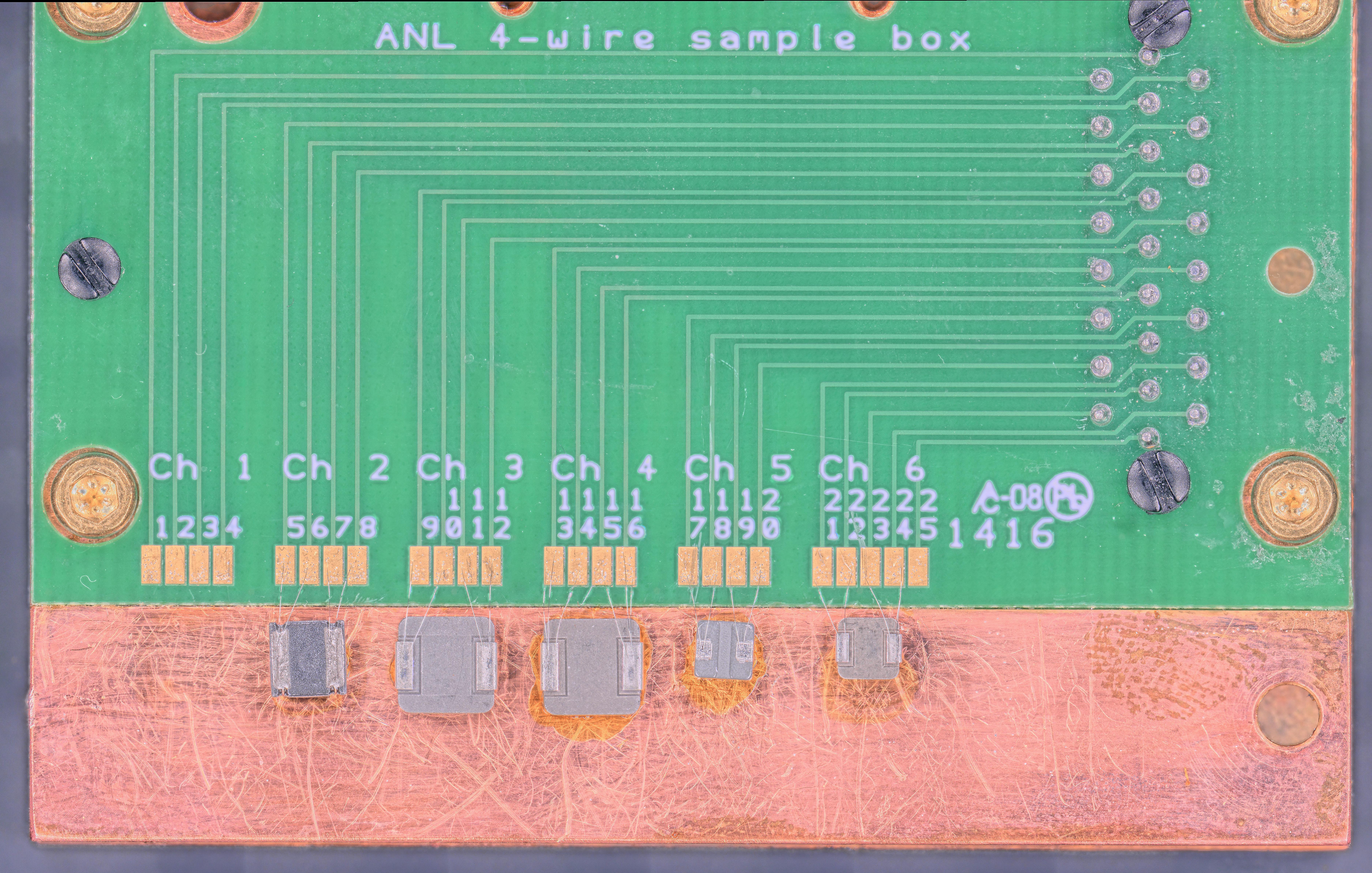}}
\caption{The five inductors mounted and wire-bonded onto the 4-wire custom testing board.}
\vspace{-10pt}
\label{cryo-inductor2}
\end{figure}

\begin{figure}[h!]
\centerline{\includegraphics[width=0.92\columnwidth]{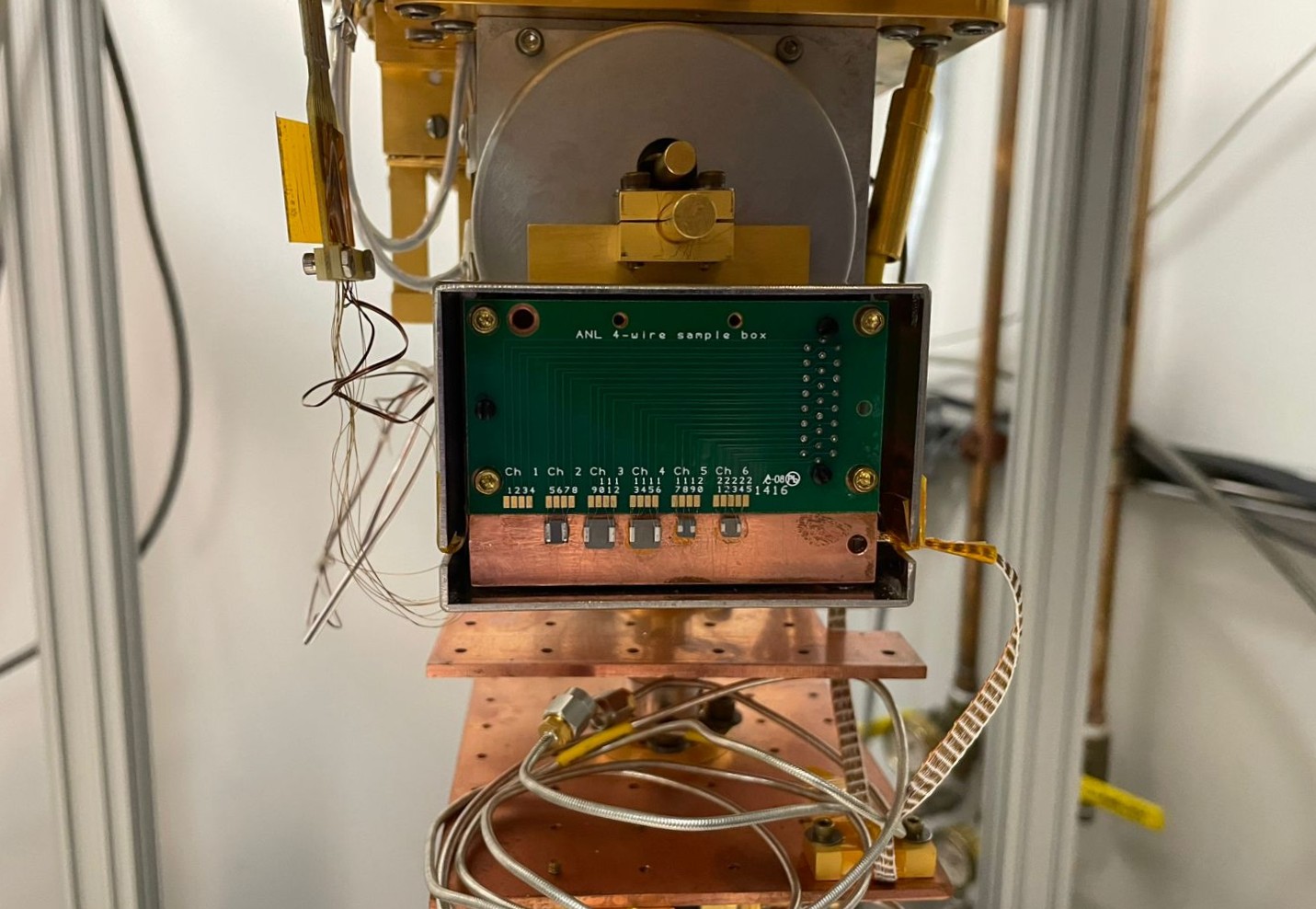}}
\caption{The 4-wire testing board positioned inside the cryostat before cooling down.}
\vspace{-10pt}
\label{board_before_cooling}
\end{figure}

\begin{figure}[h!]
\centerline{\includegraphics[width=0.92\columnwidth]{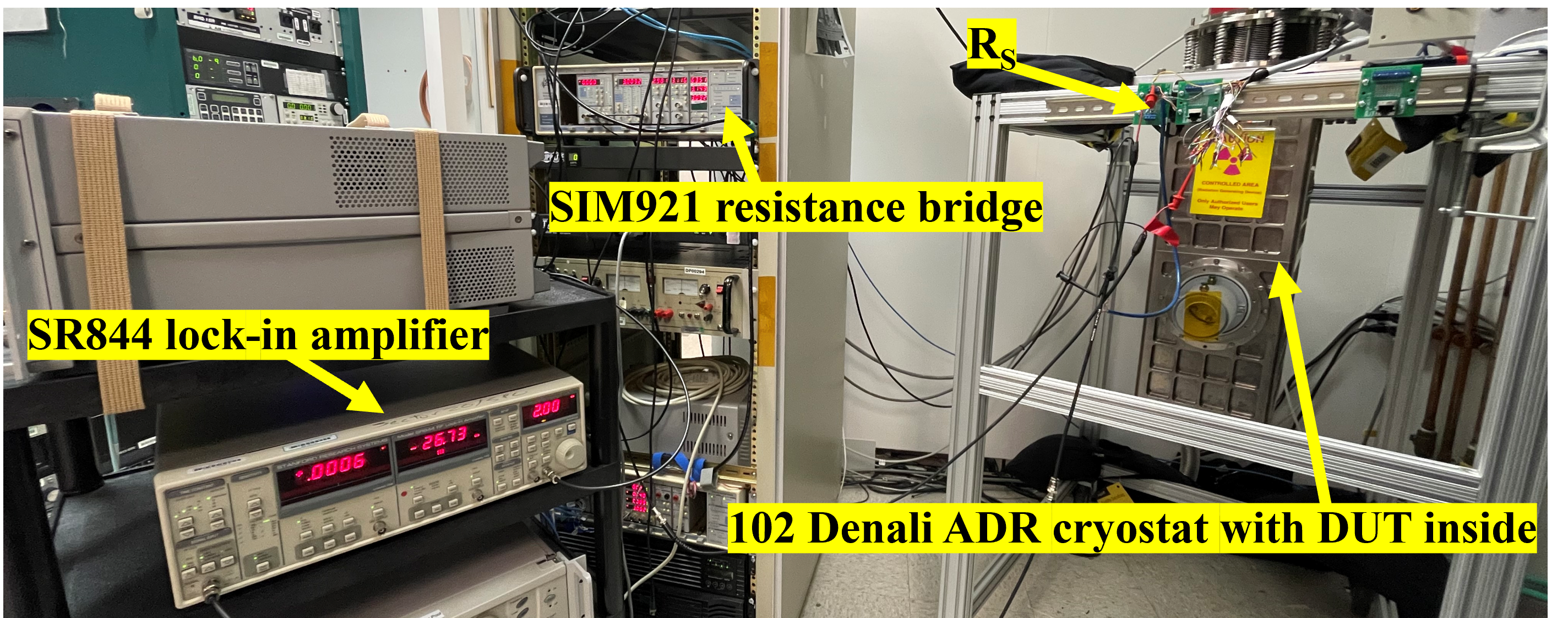}}
\caption{The measurement setup, including the cryostat, the resistance bridge, and the lock-in amplifier.}
\label{setup}
\end{figure}

Five Vishay SMD power inductors \cite{vishay_inductors} are investigated in this paper. The inductors are mounted and wire-bonded onto a custom four-wire measurement test board, as shown in Fig.~\ref{cryo-inductor2}. The nominal manufacturer-reported DC inductance and DC resistance values of the investigated inductors, together with their corresponding model names and board channel assignments, are summarized in Table~\ref{tab:inductor_models}.
A Model 102 Denali ADR cryostat \cite{denali102} was subsequently used to cool the entire test board down to 4~K. During the cooldown process, continuous $R_{dc}$ measurements of the five inductors were performed using the SIM921 resistance bridge \cite{sim921_manual}. In addition, AC inductance and AC resistance measurements as a function of frequency were carried out at both room temperature and cryogenic temperatures using the SynkTek MCL1-540 lock-in amplifier \cite{synktek_mcl1540} for frequencies up to 500~kHz and the SR844 lock-in amplifier \cite{sr844_manual} for frequencies up to 2~MHz. The assembled measurement board positioned inside the cryostat prior to cooldown is shown in Fig.~\ref{board_before_cooling}, while the complete measurement setup is shown in Fig.~\ref{setup}.

\begin{figure}[h!]
\centerline{\includegraphics[width=0.92\columnwidth]{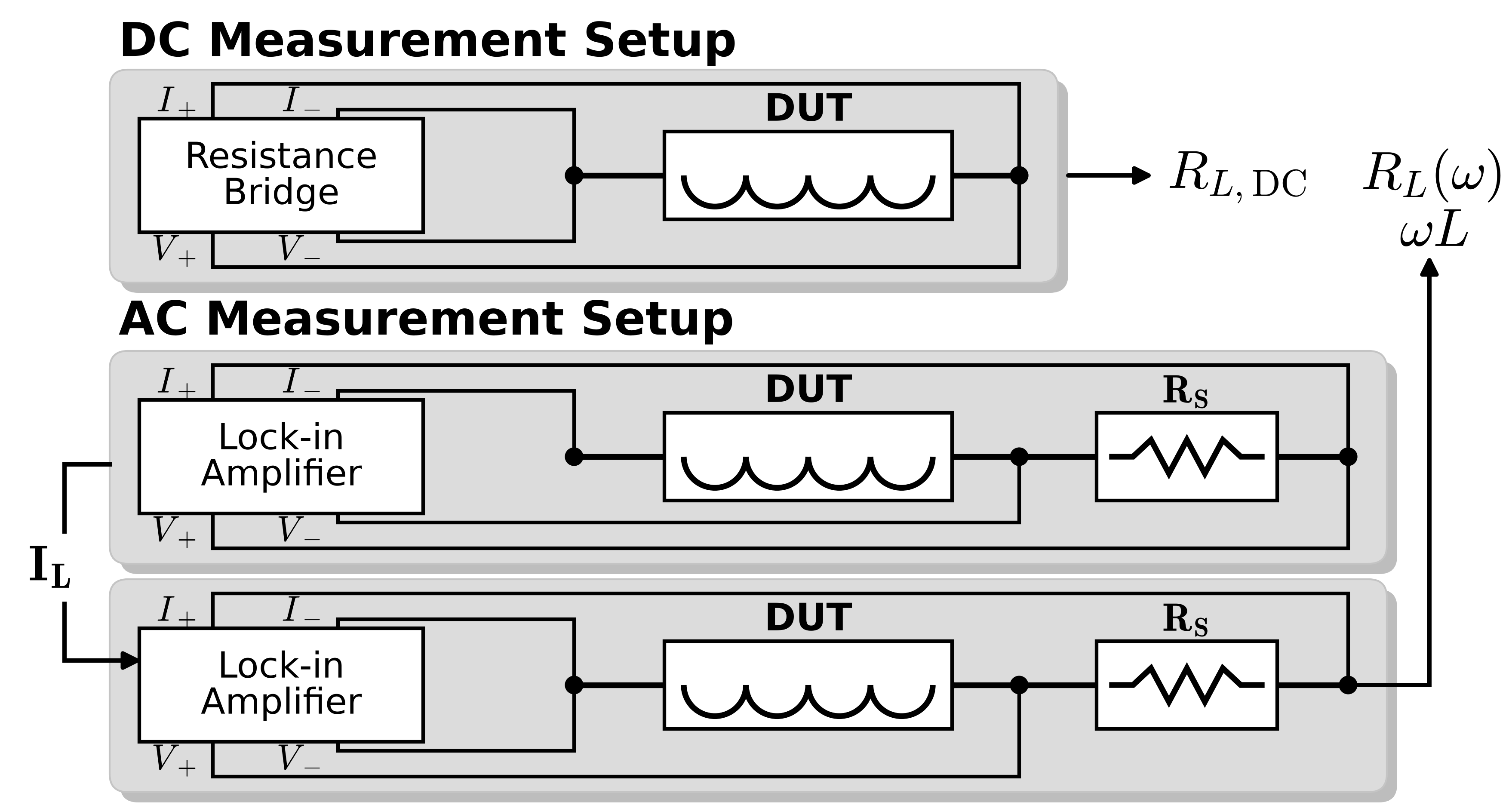}}
\caption{The measurement circuit showing DC and AC measurements.}
\vspace{-15pt}
\label{meas_setup}
\end{figure}

The measurement methodology is illustrated in Fig.~\ref{meas_setup}, in which both the DC and AC characterization configurations are included. For the DC measurements, the four wires from each inductor are directly connected to the resistance bridge. The influence of parasitic resistances associated with PCB traces, connectors, and measurement wiring is eliminated by this arrangement, whereby accurate extraction of the inductor $R_{dc}$ is allowed. Continuous $R_{dc}$ and temperature measurements were recorded throughout the cooldown process.

\begin{figure}[t]
\centerline{\includegraphics[width=\columnwidth]{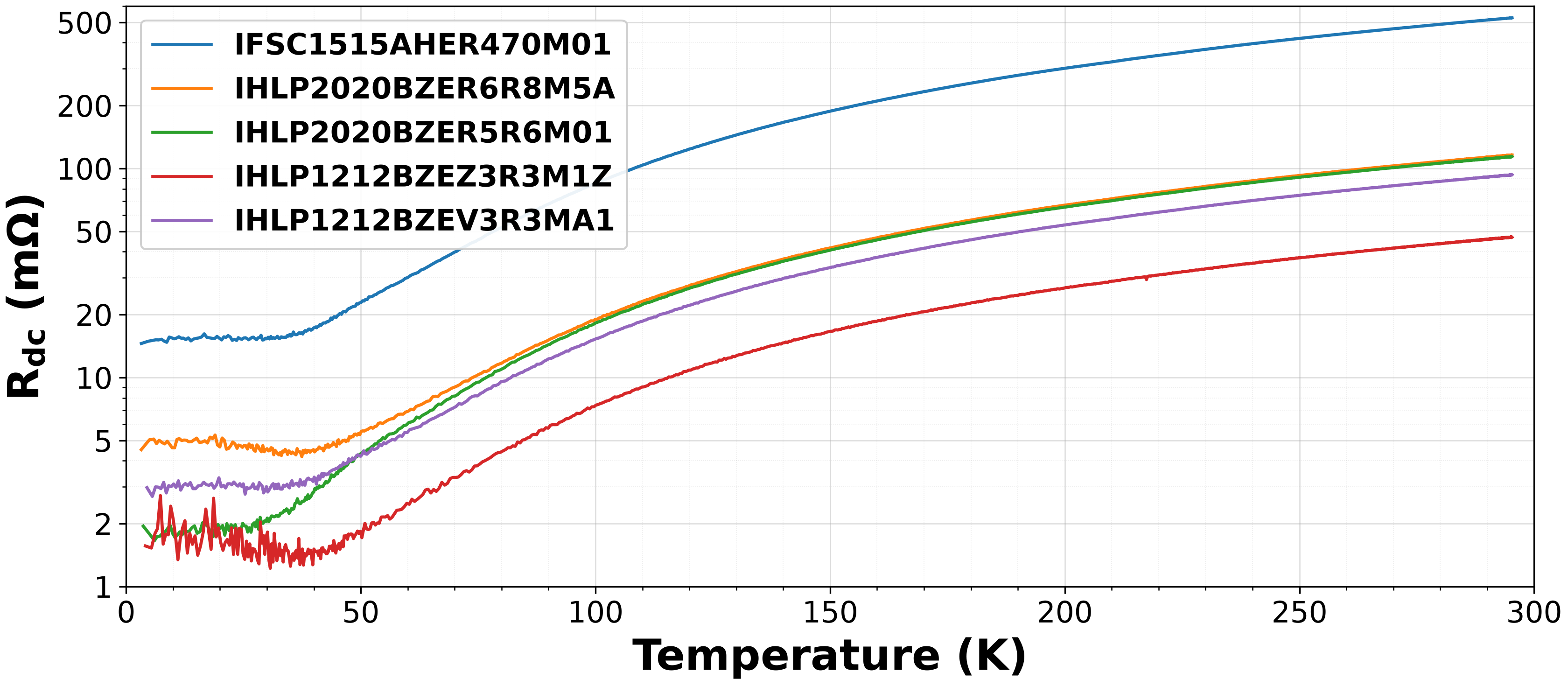}}
\caption{Measured DC resistance ($R_{dc}$) against temperature.}
\label{DCR_vs_temp}
\end{figure}
\begin{figure}[h!]
\centerline{\includegraphics[width=\columnwidth]{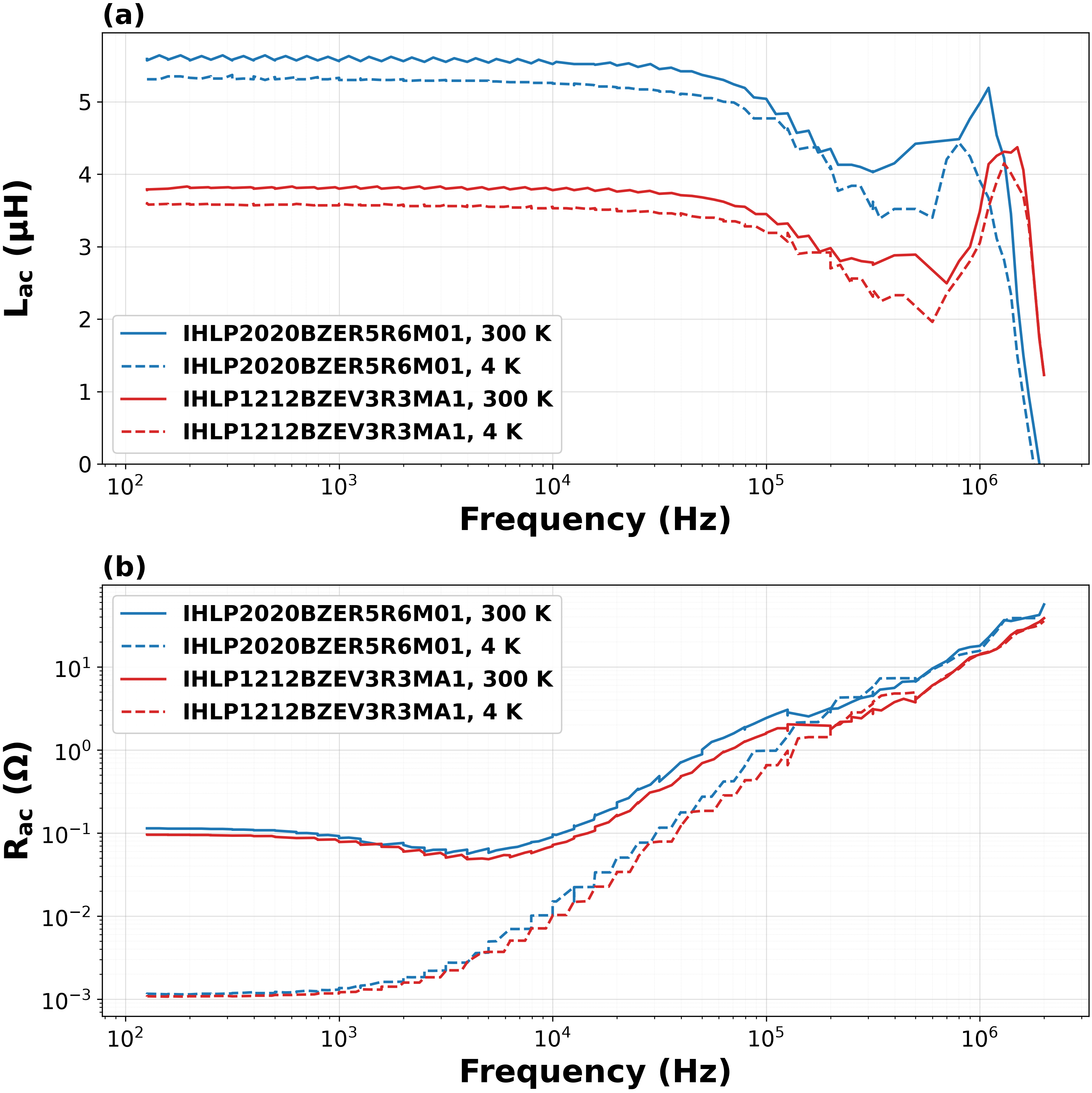}}
\caption{\rrr{Measured AC impedance at room (300\,K) and cryogenic (4\,K) temperatures, (a) inductance ($L_{ac}$), and (b) resistance ($R_{ac}$).}}
\vspace{-10pt}
\label{fig:exp_meas}
\end{figure}

For the AC characterization, a precision reference resistor, $R_s$, is connected in series with the device under test (DUT), where only the DUT is placed in 4~K. A value of 10~$\Omega$ is selected for $R_s$ to provide an impedance comparable to that of the DUT at low frequencies. Then, an unknown current is driven through the circuit using the lock-in amplifier's reference output voltage across the series connection. The voltage across the reference resistor is first measured to determine the excitation current flowing through the DUT according to
\begin{equation}
I_{DUT} = \frac{V_{Rs}}{R_s},
\label{eq:idut}
\end{equation}
where $V_{Rs}$ is the voltage measured across the reference resistor.
The complex voltage across the DUT, $V_{DUT}$, is then measured, and the complex impedance of the DUT is extracted
\begin{equation}
Z_{DUT} = \frac{V_{DUT}}{I_{DUT}}
= \frac{V_{DUT}}{V_{Rs}}R_s.
\label{eq:zdut}
\end{equation}
The measured AC resistance and inductance are, therefore
\begin{equation}
\rrr{R_{ac} = \Re\{Z_{DUT}\},}
\label{eq:rac}
\end{equation}
\begin{equation}
\rrr{L_{ac}(f) = \frac{\Im\{Z_{DUT} (f)\}}{2\pi f},}
\label{eq:lac}
\end{equation}
where $f$ is the excitation frequency, and the real and imaginary components of the measured impedance are denoted by $\Re\{\cdot\}$ and $\Im\{\cdot\}$, respectively.

\rrr{The $R_{dc}$ behavior of the five investigated inductors as a function of temperature is summarized in Fig.~\ref{DCR_vs_temp}.}
\rrr{Significant reductions in $R_{dc}$ are observed under cryogenic operation, with reduction factors ranging from approximately 25$\times$ to 58$\times$ at 4~K compared to room temperature. A noticeable settlement of the $R_{dc}$ below 30~K is observed as well.}

\begin{table}[t]
\centering
\caption{Measured Resistance and Inductance at room temperature and 4~K.}
\label{tab:all_z_reduction}
\renewcommand{\arraystretch}{1.2}
\footnotesize
\begin{tabular}{l|c|c|c}
\hline
\textbf{Model Name} & \textbf{300~K} & \textbf{4~K} & \rrr{\textbf{Factor}} \\ \hline
\multicolumn{4}{l}{\rrr{\textit{DC Resistance ($R_{dc}$)}}} \\ \hline

\scriptsize\textbf{IFSC1515AHER470M01} &
525.882~m$\Omega$ &
14.557~m$\Omega$ &
36.12$\times$\,$\downarrow$ \\ \hline

\scriptsize\textbf{IHLP2020BZER6R8M5A} &
116.083~m$\Omega$ &
4.523~m$\Omega$ &
25.67$\times$\,$\downarrow$ \\ \hline

\scriptsize\textbf{IHLP2020BZER5R6M01} &
113.944~m$\Omega$ &
1.948~m$\Omega$ &
58.49$\times$\,$\downarrow$ \\ \hline

\scriptsize\textbf{IHLP1212BZEZ3R3M1Z} &
46.936~m$\Omega$ &
1.566~m$\Omega$ &
29.97$\times$\,$\downarrow$ \\ \hline

\scriptsize\textbf{IHLP1212BZEV3R3MA1} &
93.223~m$\Omega$ &
2.978~m$\Omega$ &
31.31$\times$\,$\downarrow$ \\ \hline

\multicolumn{4}{l}{\rrr{\textit{DC Inductance ($L_{dc}$)}}} \\ \hline

\scriptsize\textbf{IHLP2020BZER5R6M01} &
5.59~\textmu H &
5.31~\textmu H &
1.053$\times$\,$\downarrow$ \\ \hline

\scriptsize\textbf{IHLP1212BZEV3R3MA1} &
3.79~\textmu H &
3.58~\textmu H &
1.059$\times$\,$\downarrow$ \\ \hline

\multicolumn{4}{l}{\rrr{\textit{AC Resistance ($R_{ac}$) @ 500~kHz}}} \\ \hline

\scriptsize\textbf{\rrr{IHLP2020BZER5R6M01}} &
\rrr{6.76~$\Omega$} &
\rrr{6.66~$\Omega$} &
\rrr{1.02$\times$\,$\downarrow$} \\ \hline

\scriptsize\textbf{\rrr{IHLP1212BZEV3R3MA1}} &
\rrr{3.75~$\Omega$} &
\rrr{4.04~$\Omega$} &
\rrr{1.08$\times$\,$\uparrow$} \\ \hline

\multicolumn{4}{l}{\rrr{\textit{AC Inductance ($L_{ac}$) @ 500~kHz}}} \\ \hline

\scriptsize\textbf{\rrr{IHLP2020BZER5R6M01}} &
\rrr{4.42~\textmu H} &
\rrr{3.52~\textmu H} &
\rrr{1.26$\times$\,$\downarrow$} \\ \hline

\scriptsize\textbf{\rrr{IHLP1212BZEV3R3MA1}} &
\rrr{2.89~\textmu H} &
\rrr{2.19~\textmu H} &
\rrr{1.32$\times$\,$\downarrow$} \\ \hline

\multicolumn{4}{l}{\rrr{\textit{AC/DC Resistance Ratio ($R_{ac}/R_{dc}$) @ 500~kHz}}} \\ \hline

\scriptsize\textbf{\rrr{IHLP2020BZER5R6M01}} &
\rrr{59.3} &
\rrr{3418.8} &
\rrr{57.7$\times$\,$\uparrow$} \\ \hline

\scriptsize\textbf{\rrr{IHLP1212BZEV3R3MA1}} &
\rrr{40.22} &
\rrr{1356.61} &
\rrr{33.7$\times$\,$\uparrow$} \\ \hline

\end{tabular}
\end{table}

AC measurements up to 2~MHz are performed for only two inductors, namely \textit{IHLP2020BZER5R6M01} and \textit{IHLP1212BZEV3R3MA1}, since the calculated switching frequencies for these devices \rrr{range from approximately 360~kHz to 670~kHz} (see Section~\ref{Performance_Comparison}) 
and a 2~MHz sweep is sufficient for the first several harmonics above the target switching frequency to be captured. The corresponding measurement results are summarized in Fig.~\ref{fig:exp_meas}. All measured impedances are summarized in Table~\ref{tab:all_z_reduction}.

Both the frequency-dependent winding resistance and the magnetic core loss contribution are included in the measured AC resistance, with the latter becoming increasingly dominant at higher frequencies while remaining relatively negligible at lower frequencies. It can be observed that the reduction in DC inductance at 4~K is relatively small (approximately 5--5.5\%) compared to the substantial reduction in DC resistance under cryogenic operation. However, as discussed previously, the ratio between AC and DC resistance increases dramatically at cryogenic temperatures compared to room-temperature operation. This behavior is primarily attributed to the increased core losses resulting from the reduced core resistivity, in addition to the steeper increase in AC winding resistance with frequency caused by the intensified skin effect at low temperatures.

\section{Performance Comparison}
\label{Performance_Comparison}

\begin{figure}[t]
\centerline{\includegraphics[width=\columnwidth]{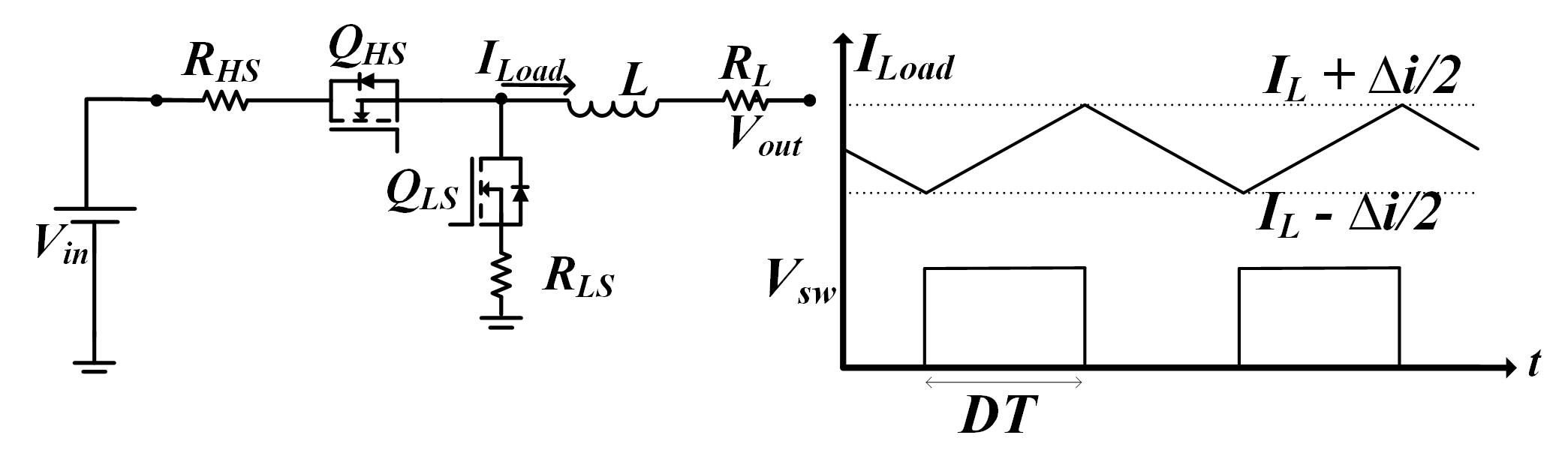}}
\caption{A single-phase buck converter and its inductor current waveform.}
\label{buck_conv}
\end{figure}
\begin{table}[t]
\centering
\caption{\rrr{Calculated Switching Frequencies at 1~A, 2~A, and 3~A Current Loads.}}
\label{tab:fsw_comparison}
\renewcommand{\arraystretch}{1.2}
\setlength{\tabcolsep}{5pt}
\footnotesize
\begin{tabular}{l|c|c|c|c}
\hline
\textbf{Model Name} & \textbf{$T$} & \rrr{\textbf{1~A}} & \rrr{\textbf{2~A}} & \rrr{\textbf{3~A}} \\ \hline
\textbf{IHLP2020BZER5R6M01} & \rrr{300~K} & \rrr{387 kHz} & \rrr{430 kHz} & \rrr{472 kHz} \\
& \rrr{4~K}   & \rrr{363 kHz} & \rrr{365 kHz} & \rrr{367 kHz} \\ \hline
\textbf{IHLP1212BZEV3R3MA1} & \rrr{300~K} & \rrr{560 kHz} & \rrr{614 kHz} & \rrr{667 kHz} \\
& \rrr{4~K}   & \rrr{539 kHz} & \rrr{543 kHz} & \rrr{546 kHz} \\ \hline
\end{tabular}

\end{table}

To evaluate the effectiveness of operating these inductors under cryogenic conditions, their behavior and loss characteristics are analyzed \rrr{within a single-phase 24~V/1~V buck converter at load currents of 1~A, 2~A, and 3~A with a 0.5~A peak-to-peak current ripple.} The topology of the converter is shown in Fig.~\ref{buck_conv}. The focus is placed on investigating the inductor-related losses and the corresponding efficiency behavior under cryogenic operation. 




The analysis is performed for the two inductors selected for the AC characterization measurements. For each inductor, the required switching frequency is calculated at both the room and 4~K temperatures, 
according to

\begin{equation}
\rrr{D = \frac{V_{out} + I_{Load}(R_{dc} + R_{LS,on})}{V_{in} - I_{Load}(R_{HS,on} - R_{LS,on})},}
\label{eq:duty_cycle}
\end{equation}

\begin{equation}
f_{sw} =
\frac{D \left[V_{in} - I_{Load}(R_{HS,on} + R_L) - V_{out}\right]}
{L\Delta i},
\label{eq:ripple_current}
\end{equation}
where $D$ is the converter duty cycle, $V_{in}$ is the input voltage, $V_{out}$ is the output voltage, $I_{Load}$ is the load current, $R_{HS,on}$ is the on-state resistance of the high-side switch, $R_{LS,on}$ is the on-state resistance of the low-side switch, $R_L$ is the DC resistance of the inductor, $f_{sw}$ is the converter switching frequency, $L$ is the inductance value, and $\Delta i$ is the peak-to-peak inductor current ripple. 
\rrr{For this analysis, the high-side and low-side switches are the commercial GaN FETs EPC2040 and EPC2216, respectively, and their on-state resistances are treated as temperature dependent. At room temperature, $R_{HS,on}$ and $R_{LS,on}$ are set to their typical datasheet values of 23.857~m$\Omega$ (EPC2040) and 19.924~m$\Omega$ (EPC2216)~\cite{epc_ganfets}. 
Based on the experimentally measured cryogenic behavior of a comparable EPC eGaN device~\cite{colmenares2016experimental}, an 80\% reduction in on-state resistances is assumed for the EPC2040 and EPC2216 at 4~K. Accordingly, their room-temperature on-state resistances are reduced to 4.771 and 3.985~m$\Omega$, respectively. This literature-based scaling provides a first-order representation of cryogenic switch conduction loss; other temperature-dependent device parameters are retained at their room-temperature values.
}
The resulting switching frequencies are summarized in Table~\ref{tab:fsw_comparison}. Although the inductance slightly decreases at 4~K for these inductors ($\sim$5\% as shown in Table~\ref{tab:all_z_reduction}), the required switching frequency is also reduced under cryogenic operation. Thus, (\ref{eq:ripple_current}) remains near-optimal across the temperature range. This behavior is primarily attributed to the substantial reduction in the inductor DC resistance at 4~K, by which the relatively minor decrease in inductance is outweighed.

\begin{figure}[tb]
\centerline{\includegraphics[width=\columnwidth]{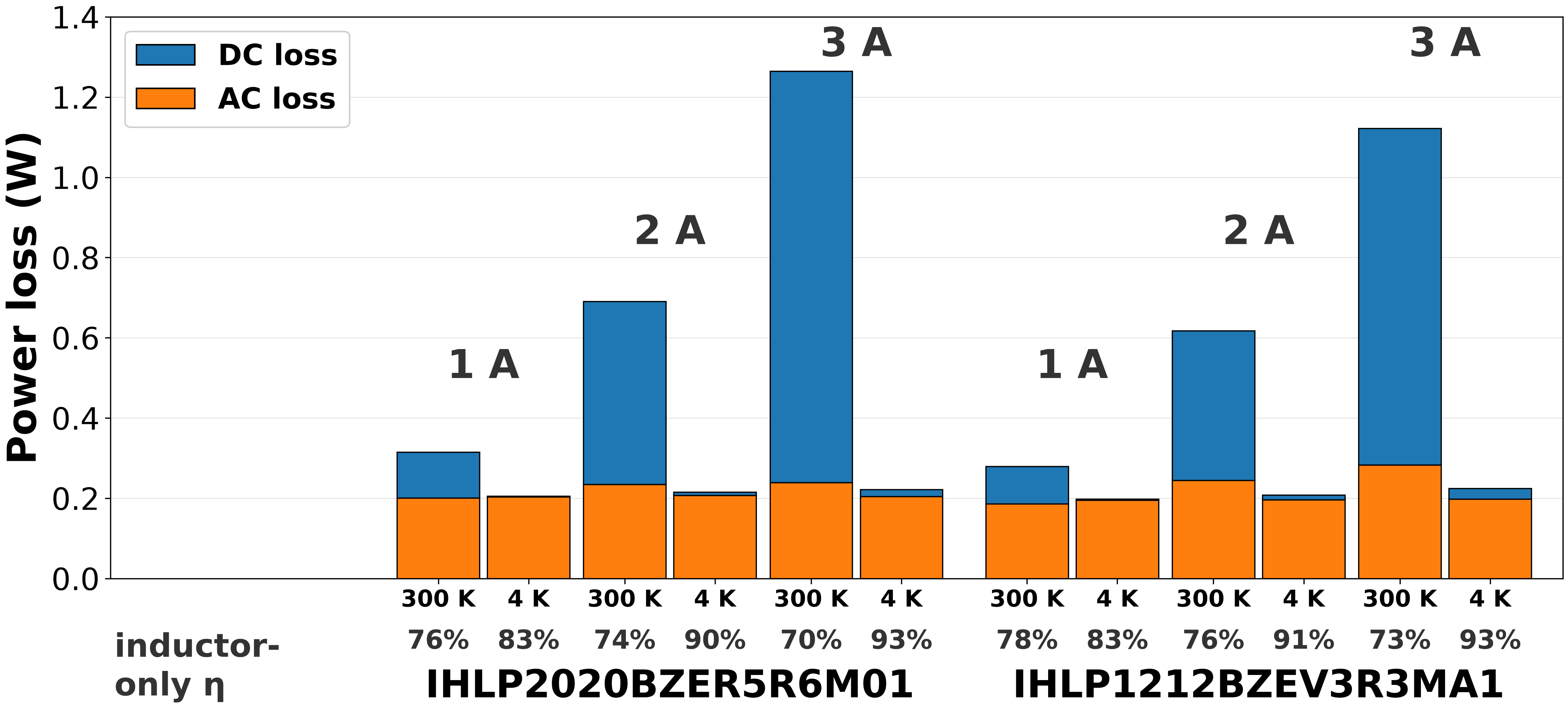}}
\caption{Analytical power-loss breakdown (stacked DC and AC loss, with the total
labelled on top of each bar) and efficiency (listed below each bar) of the
IHLP2020BZER5R6M01 and IHLP1212BZEV3R3MA1 inductors at 300~K and
4~K.}
\label{fig:loss_efficiency}
\end{figure}

The loss and efficiency of the inductor in each case are calculated as \cite{ramiectc},
\begin{equation}
P_{dc} = I_{L,dc}^2  R_{L,dc},
\label{eq:loss_dc}
\end{equation}
\begin{equation}
P_{ac} = \frac{2\Delta i^2}{D^2(1-D)^2} \sum_{k=1}^{N} \frac{\sin^2(k\pi D)}{(k\pi)^4} R_{L,ac}(2k\pi f_{sw}),
\label{eq:loss_ac}
\end{equation}
\begin{equation}
\eta = \frac{I_{Load}V_{out}}{I_{Load}V_{out}+P_{loss}},
\end{equation}
where 
$I_{L,dc}$ is the DC component of the inductor current, and $R_{L,dc}$ is the DC resistance of the inductor winding, 
$N$ is the number of considered harmonic components, $k$ is the harmonic index, and $R_{L,ac}$ is the AC resistance at this specific frequency harmonic. Note that $R_{L,ac}(2k\pi f_{sw})$ is the frequency-dependent AC resistance of the inductor evaluated at the $k^{\mathrm{th}}$ harmonic frequency. The resulting losses and efficiencies are shown in Fig.~\ref{fig:loss_efficiency}.

\rrr{A significant reduction in DC conduction loss is observed at 4~K for all load currents, directly following the reduction in winding DC resistance reported in Table~\ref{tab:all_z_reduction} through (\ref{eq:loss_dc}). The resulting DC loss is reduced by 58.49$\times$ for theIHLP2020BZER5R6M01 and 31.31$\times$ for the IHLP1212BZEV3R3MA1. The AC loss, in contrast, is affected by cooling in a load-dependent manner. At the 1~A load, the AC loss is slightly \emph{increased} by cryogenic operation (from 200.9 to 203.2~mW for the IHLP2020BZER5R6M01 and from 185.8 to 195.2~mW for the IHLP1212BZEV3R3MA1), whereas at 2~A and 3~A it is \emph{decreased} at 4~K. This cross-over is explained by the load-dependent switching frequency at each temperature. At 4~K, the winding $R_{dc}$ is negligible, and the switching frequency required to maintain a constant ripple of $\sim$0.5~A, as computed from (\ref{eq:ripple_current}), remains nearly independent of load. Consequently, the AC loss, as computed from (\ref{eq:loss_ac}), also remains nearly constant for both inductors (see Fig.~\ref{fig:loss_efficiency}). 
In contrast, at 300~K, the switching frequency is forced to increase with load by the large winding $R_{dc}$ (e.g., 387$\rightarrow$430$\rightarrow$472~kHz for the IHLP2020BZER5R6M01), pushing the current harmonics toward higher-$R_{ac}$ frequencies and the 300~K AC loss is increased with load (201$\rightarrow$235$\rightarrow$239~mW). At 1~A, the 300~K switching frequency remains sufficiently low that the higher frequency-dependent resistance at 4~K slightly increases the AC loss. 
At 2~A and above, the load-driven increase in the 300~K switching frequency raises the 300~K AC loss above the nearly constant 4~K value, resulting in a net AC-loss reduction upon cooling. 
Because the DC-loss reduction is dominant in every case, the inductor-only efficiency is improved at 4~K for all loads, with the improvement increasing with load: 
approximately 6.9, 16.0, and 22.8 percentage points at 1, 2, and 3~A for the IHLP2020BZER5R6M01, and 5.3, 14.2, and 20.2 percentage points for the IHLP1212BZEV3R3MA1.}

\begin{figure}[tb]
\centerline{\includegraphics[width=0.9\columnwidth]{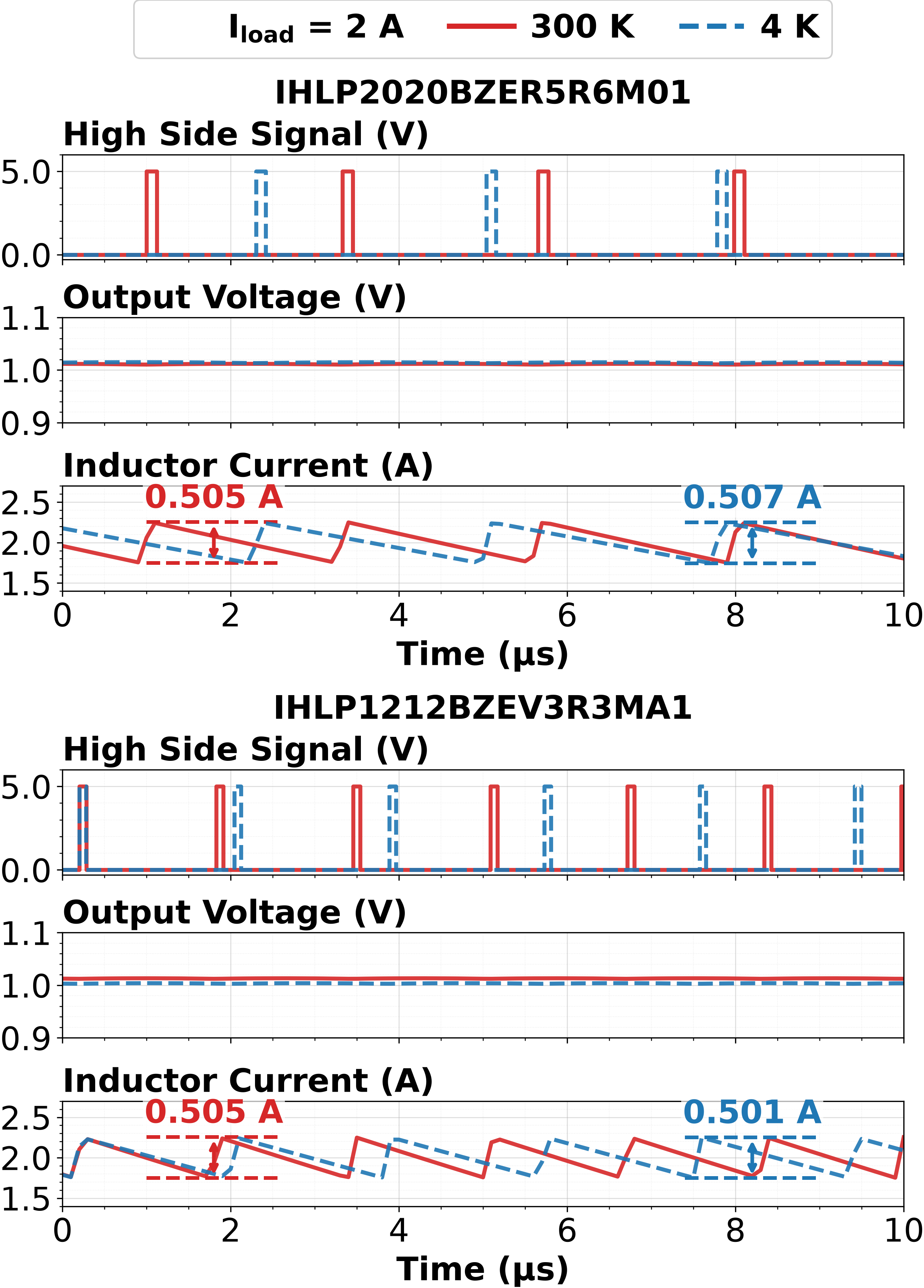}}
\caption{\rrr{Simulated buck converter performance for the IHLP2020BZER5R6M01 (top) and IHLP1212BZEV3R3MA1 (bottom) at 2\,A load at 300\,K (red) and 4\,K (blue).}}
\vspace{-10pt}
\label{waveforms_6box}
\end{figure}

Although the primary motivation for placing the power converter inside the cryogenic environment is not to improve converter efficiency by itself, but rather to eliminate the conduction losses associated with the long interconnects between an external power converter and the cryogenic electronics, an additional benefit is demonstrated by the results. Besides the reduction of interconnect-related conduction losses, improved overall efficiency under cryogenic operation is also exhibited by the investigated inductors. Consequently, a dual advantage is provided by the integration of the power converter within the cryogenic environment: reduced power delivery losses and improved magnetic component efficiency.

To ensure that self-heating does not raise the inductor temperature outside the cryogenic operating regime, the following first-order thermal analysis is considered. The calculated inductor loss at 4~K remains approximately 0.2--0.25~W over the considered operating range. As shown in Fig.~\ref{fig:loss_efficiency}, the measured winding resistance remains near its minimum below approximately 20--30~K. Using the first-order relation $R_{\mathrm{th,eff}}=(T_L-T_{\mathrm{stage}})/P_{\mathrm{loss}}$, where $T_L$ is the inductor temperature and $T_{\mathrm{stage}}=4$~K is the cryogenic stage temperature, maintaining the inductor within this low-resistance regime requires an effective thermal resistance to the cryogenic stage of approximately 60--130~K/W, depending on the operating point and selected temperature limit. This requirement provides substantial thermal-design margin relative to values experimentally demonstrated with dedicated cryogenic packaging; for example, chip-to-cold-stage thermal resistances of approximately 1--2~K/W have been reported near 4~K using high-conductivity package materials and engineered thermal interfaces~\cite{durandetto2019thermal}. Although the thermal path of an SMD inductor differs from that implementation, the comparison indicates that the required 60--130~K/W range is compatible with practical cryogenic packaging with appropriate thermal coupling to the cold stage. The actual thermal resistance remains implementation dependent and is determined by the inductor package, board, interfaces, and coupling to the cryogenic stage; detailed electrothermal characterization is outside the scope of this work.

\subsection{\rrr{Transient Converter-Level Simulation}}
\rrr{To verify the complete converter waveforms and the interaction between the frequency-dependent inductor loss and the switching harmonics, the full single-phase 24~V-to-1~V buck converter is simulated in the time domain in Cadence Spectre for each inductor at both temperatures and at load currents of 1, 2, and 3~A. The measured frequency-dependent impedance up to 2~MHz is imported as a Touchstone one-port and incorporated into the buck converter for the corresponding temperature (Table~\ref{tab:all_z_reduction}). Because both the frequency-dependent winding resistance and the magnetic core loss are already embedded in the measured AC resistance, the true harmonic loading of the inductor is reproduced by this representation without a separate core-loss model being required. 
The GaN power devices, EPC2040 and EPC2216, are re-modeled at cryogenic temperature and switched at the preferred  frequency computed for each case (Table~\ref{tab:fsw_comparison}).}

\begin{figure}[t]
\vspace{-15pt}
\centerline{\includegraphics[width=\columnwidth]{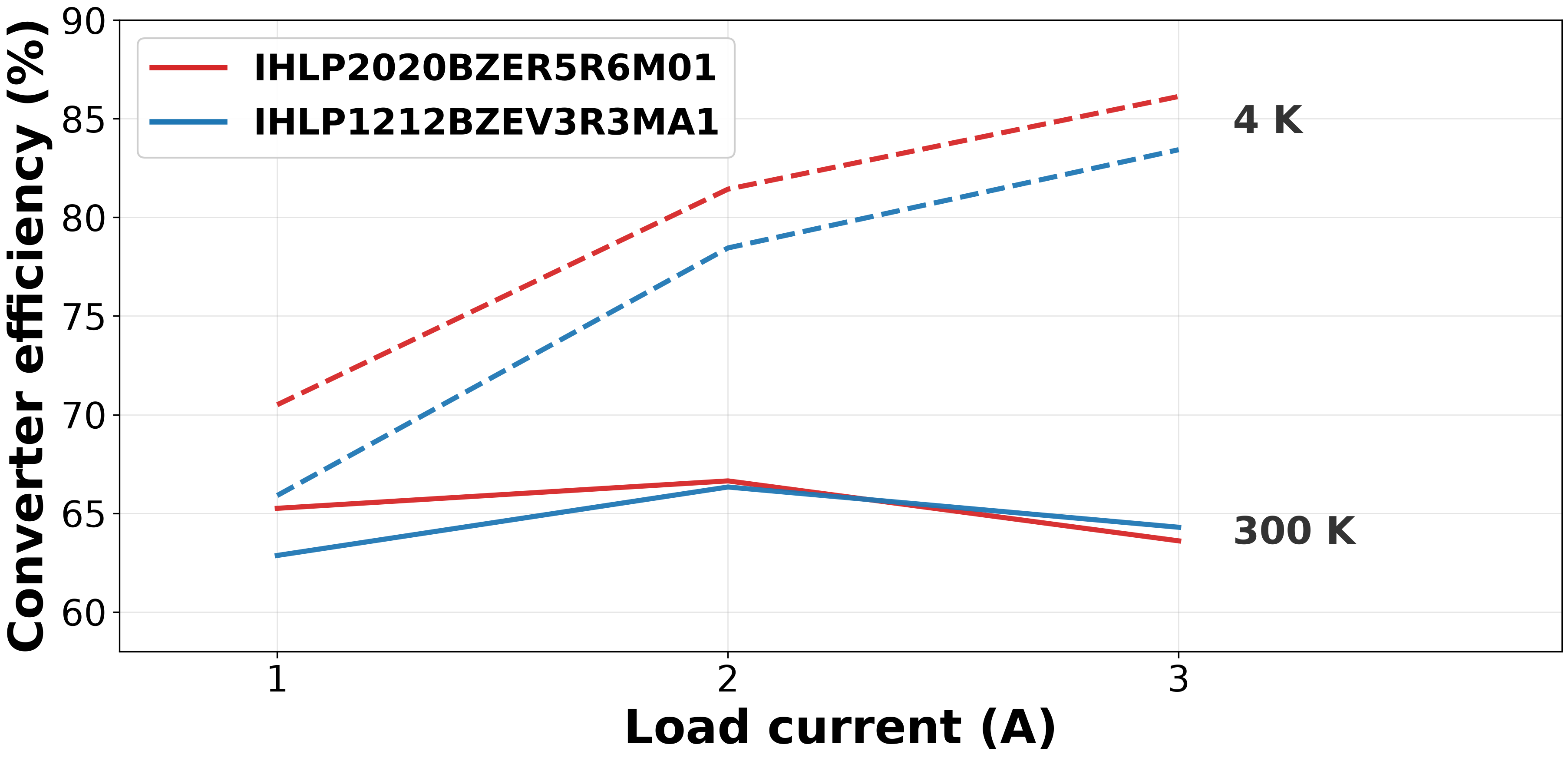}}
\caption{\rrr{Simulated full-converter efficiency versus load current for the two AC-characterized inductors at 300~K and 4~K.}}
\vspace{-5pt}
\label{fig:eff_vs_current}
\end{figure}

\rrr{The resulting steady-state waveforms at the representative 2~A load are shown in Fig.~\ref{waveforms_6box}. For both inductors and at both temperatures, the output voltage is tightly regulated at 1~V and the ripple is within the 0.5~A target.} 
\rrr{The full-converter efficiencies extracted from these transient simulations are plotted against load current in Fig.~\ref{fig:eff_vs_current}. 
At 300~K, the efficiency peaks near 2~A and then falls (e.g., 65.25\%\,$\rightarrow$\,66.64\%\,$\rightarrow$\,63.60\% for the IHLP2020BZER5R6M01), as the higher winding DC resistance causes the $I^2 R_{dc}$ conduction loss to grow rapidly with load. At 4~K, by contrast, the efficiency increases monotonically with load---from 70.50\% at 1~A to 81.43\% at 2~A and 86.13\% at 3~A for the IHLP2020BZER5R6M01, and from 65.90\% to 78.45\% and 83.43\% for the IHLP1212BZEV3R3MA1. This trend results from the substantial reduction in winding DC resistance, which renders the DC conduction loss negligible, while the nearly load-independent AC and switching losses are amortized over an increasing output power.
Consequently the cryogenic efficiency benefit widens with load, from 3.0--5.2 percentage points at 1~A to 19.1--22.5 percentage points at 3~A. While the converter-level efficiencies are lower than the inductor-only values in Fig.~\ref{fig:loss_efficiency}, the cryogenic improvement persists at the converter level, confirming that the reduction in winding DC resistance remains the dominant factor even when switch losses are considered.}

\section{Conclusion}
\label{Conclusion}
\rrr{Experimental characterization of commercially available Vishay SMD power inductors is presented in this paper under cryogenic operation down to 4~K, including continuous $R_{dc}$ measurements and frequency-dependent AC characterization. Substantial reductions in DC resistance of approximately 25$\times$--58$\times$ are demonstrated relative to room temperature with only slight inductance degradation of 5--5.5\%. While the AC resistance and core losses are shown to increase with frequency at 4~K, the overall loss behavior is dominated by the reduced DC conduction loss.}

\rrr{A representative single-phase buck converter is designed with the experimentally characterized inductors and GaN power switches modeled at 4~K using measured cryogenic on-resistance data. The converter is simulated in Spectre, exhibiting inductor-only efficiency gains of 5.3--6.9 percentage points at 1~A and up to 20.2--22.8 percentage points at 3~A. 
The full-converter efficiency is also shown to improve by 3.0--5.2 percentage points at 1~A and by up to 22.5 percentage points at 3~A, highlighting the advantage of cryogenic power conversion at higher current loads. To the best of authors' knowledge, this is the first converter-level characterization of commercial off-the-shelf SMD power inductors down to 4~K for cryogenic DC--DC power conversion. These results demonstrate that off-the-shelf power inductors offer a practical, cost-effective solution for cryogenic power delivery, improving inductor efficiency while also reducing the conduction losses of long power-delivery interconnects.}

\section*{Acknowledgment}
This work was supported by the Center for Heterogeneous Integration of Micro Electronic Systems (CHIMES), one of seven centers in Joint University Microelectronics Program (JUMP) 2.0, a Semiconductor Research Corporation (SRC) program sponsored by the Defense Advance Research Project Agency (DARPA).

This work was supported by the U.S. Department of Energy, Office of Science, Advanced Scientific Computing Research (ASCR) and Basic Energy Sciences (BES), under Contract No. DE-AC02-06CH11357 at Argonne and DE-AC02-76SF00515 at SLAC. This work utilized resources at the Advanced Photon Source, a U.S. Department of Energy Office of Science User Facility. Work performed at the Center for Nanoscale Materials and Advanced Photon Source, both U.S. Department of Energy Office of Science User Facilities (SUF), was supported by the U.S. DOE, Office of Basic Energy Sciences, under Contract No. DE-AC02-06CH11357. Work at University of Chicago is supported by the Divisions of Chemistry (CHE) and Materials Research (DMR), National Science Foundation, under grant numbers NSF/CHE-1834750 and NSF/CHE-2335833. This work was primarily supported by the AUREIS project (part of Microelectronics Energy Efficiency Research Center for Advanced Technologies (MEERCAT)) and the Morpheus project, supported by DOE BES/SUF's Accelerator and Detector R\&D program.

\bibliographystyle{IEEEtran}
\bibliography{Refs}
\end{document}